\documentclass[10pt,letterpaper,conference]{IEEEtran}
\IEEEoverridecommandlockouts
\usepackage{amsmath,amssymb,amsfonts}
\usepackage{gensymb}
\usepackage{array}
\usepackage{algorithmic}
\usepackage{graphicx}
\usepackage{textcomp}
\usepackage{xcolor}
\usepackage{xspace}
\usepackage{enumitem}
\usepackage{url}
\usepackage{breakurl}
\usepackage{caption}
\usepackage{subcaption}
\usepackage{multirow}
\usepackage[normalem]{ulem}
\usepackage[numbers,sort&compress]{natbib}
\usepackage[linesnumbered,ruled,lined,boxed]{algorithm2e}
\graphicspath{{figures/}}
\SetKwInput{KwInput}{Input}                
\SetKwInput{KwOutput}{Output}              

\newcolumntype{L}[1]{>{\raggedright\let\newline\\\arraybackslash\hspace{0pt}}m{#1}}
\newcolumntype{C}[1]{>{\centering\let\newline\\\arraybackslash\hspace{0pt}}m{#1}}
\newcolumntype{R}[1]{>{\raggedleft\let\newline\\\arraybackslash\hspace{0pt}}m{#1}}

\newcommand{\coloroverride}[2]{\textcolor{#1}{#2}}
\renewcommand{\coloroverride}[2]{\textcolor{black}{#2}}

\newcommand{\chk}[1]{\coloroverride{red}{#1}} 

\newcommand{\ie}{\textit{i.e.,}\xspace}
\newcommand{\eg}{\textit{e.g.,}\xspace}

\newcommand{\att}{MNO-A\xspace}
\newcommand{\tmo}{MNO-B\xspace}

\begin{document}

\title{Neutral-Hosts In The Shared Mid-Bands: Addressing Indoor Cellular Performance
}

\author{
\IEEEauthorblockN{
Muhammad Iqbal Rochman\IEEEauthorrefmark{1},
Joshua Roy Palathinkal\IEEEauthorrefmark{1},
Vanlin Sathya\IEEEauthorrefmark{2},
Mehmet Yavuz\IEEEauthorrefmark{2},
and Monisha Ghosh\IEEEauthorrefmark{1}}
\IEEEauthorblockA{
\IEEEauthorrefmark{1}University of Notre Dame, 
\IEEEauthorrefmark{2}Celona, Inc.\\ 
Email: \IEEEauthorrefmark{1}\{mrochman,jpalathi,mghosh3\}@uchicago.edu, \IEEEauthorrefmark{2}\{vanlin,mehmet\}@celona.io}
}



\maketitle


\begin{abstract}

The 3.55  - 3.7 GHz Citizens Broadband Radio Service (CBRS) band in the U.S., shared with incumbent Navy radars, is witnessing increasing deployments both indoors and outdoors using a shared, licensed model. Among the many use-cases of such private networks is the indoor neutral-host, where cellular customers of Mobile Network Operators (MNOs) can be seamlessly served indoors over CBRS with improved performance, since building loss reduces the indoor signal strength of mid-band 5G cellular signals considerably. In this paper, we present the first detailed measurements and analyses of a real-world deployment of an indoor private network serving as a neutral-host in the CBRS band serving two MNOs. Our findings demonstrate significant advantages:
(i) minimal outdoor interference from the CBRS network due to over 22 dB median penetration loss, ensuring compatibility with incumbent users; (ii) substantial indoor performance gains with up to 535\(\times\) and 33\(\times\) median downlink and uplink throughput improvements, respectively, compared to the worst-performing MNO; (iii) reduced uplink transmit power for user devices (median 12 dB reduction), increasing energy efficiency; and (iv) significant capacity offload from the MNO network (median 233 resource blocks/slot freed in 5G), allowing MNOs to better serve outdoor users. These results highlight the potential of low-power indoor CBRS deployments to improve performance, increase spectrum efficiency, and support coexistence with current and future incumbents, \eg the 3.1 - 3.45 GHz band being considered for sharing with federal incumbents in the U.S.


\end{abstract}



\section{Introduction}\label{introduction}

Most cellular networks today are deployed by public Mobile Network Operators (MNOs) over high-power, exclusively licensed spectrum and primarily serve mobile consumers through smartphone connections. However, a recent report indicates a slowing growth-rate in mobile data usage~\cite{ericsson2024mobility}, while also noting that about 80\% of mobile data usage occurs indoors, highlighting the need to reexamine the need for a continued increase of traditional, high-powered, exclusively licensed spectrum that is best suited to provide outdoor mobile coverage. On the other hand, a recent report from the National Telecommunications and Information Administration (NTIA) in the U.S.~\cite{ntia2024cbrs} documents the rise in private networks deployed over the shared 3.55  - 3.7 GHz Citizens Broadband Radio Service (CBRS) band in the U.S., serving many diverse use cases, both indoors and outdoors, such as neutral-host networks, warehouse and factory connectivity, remote oil-field monitoring etc. These private networks operate at low (30 dBm/10 MHz) and medium (47 dBm/10 MHz) Effective Isotropic Radiated Power (EIRP) power levels in order to enable coexistence with both incumbent users of the band (Navy radar) as well as with shared licensees using either Priority Access License (PAL) or General Authorized Access (GAA). Similary, Ofcom in the U.K. is seeking to increase locally licensed shared spectrum usage in 3.8 - 4.2 GHz in order to satisfy the growing needs of private networks~\cite{ofcom2024_3GHz}.

\noindent{\bf{Indoor-connectivity Challenges:}} Historically, in-building connectivity to public MNOs has been a complex challenge, exacerbated by the advent of energy-efficient buildings that feature low-E glass windows that increase RF attenuation especially at the mid-band frequencies favored by 5G. This is especially true for uplink connections to an outdoor base-station (BS), since mobile phones are limited in their transmit power capabilities. Existing solutions, such as Distributed Antenna Systems (DAS), offer improved indoor coverage but come with high costs and limited scalability. A DAS amplifies cellular signals through extensive passive cabling installed throughout a building. This intricate system requires an expensive design, specialized infrastructure, and the installation of RF antennas to manage and transmit signals.


\begin{figure}
    \includegraphics[width=\linewidth]{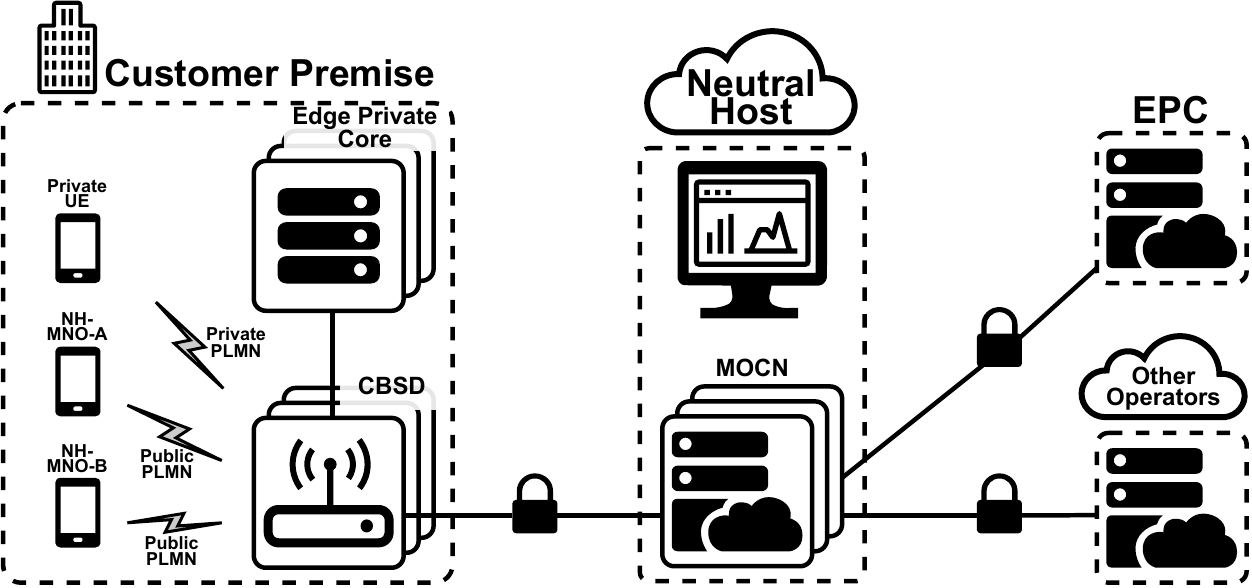}
    \caption{Neutral-host architecture.}
    \label{fig:nh_architecture_2}
    \vspace{-1em}
\end{figure}


\noindent{\bf{Neutral-host networks:}} Neutral-host networking over the CBRS band in the U.S. has emerged as a promising alternative to DAS~\cite{sathya2023nh}. A neutral-host network leverages an organization's private 4G/5G wireless network infrastructure to securely deliver third-party wholesale carrier services to users wishing to access public networks. This is achieved by routing packets between networks using the Multi-Operator Core Network (MOCN) architecture~\cite{3gpp23251}, as shown in Fig.~\ref{fig:nh_architecture_2}. By allowing multiple MNOs to share a single private network, a neutral-host network enhances connectivity for mobile subscribers and extends cellular service coverage. One of the key benefits is the ability to improve signal quality and increase capacity in areas where cellular coverage is typically weak, such as inside large buildings. Moreover, it integrates smoothly with an MNO's existing network, requiring no user actions or authentication. Users can connect automatically and seamlessly through the neutral-host, as their devices detect the network by scanning the CBRS frequency band for identifiable signals. When a device identifies a neutral-host, it retrieves a list of Public Land Mobile Network (PLMN) IDs from the broadcast network signature. If there is a commercial roaming agreement between the mobile operator and the neutral-host, the neutral-host broadcasts the relevant MNO's PLMN-ID, allowing the user to securely access their data services through a tunnel between the neutral-host and the MNO using the SIM credentials associated with the user's plan. This solution reduces infrastructure costs and allows MNOs to manage traffic more efficiently, alleviating congestion in high-demand situations while providing scalable support to meet evolving user needs.

\noindent{\bf{Spectrum Challenges:}} The key requirement for such neutral-host networks is access to sufficient spectrum. Since these networks are mostly deployed indoors, an EIRP of 30 dBm/10 MHz, as permitted by CBRS, is sufficient to meet coverage and capacity requirements. Such low-power, indoor use is also an attractive solution to the spectrum sharing challenges in many mid-band frequencies, such as the 3.1 - 3.45 GHz and 7.125 - 8.4 GHz bands where there are multiple federal incumbents which operate outdoors, and it is a challenge to either relocate them or develop coexistence schemes for high-power, outdoor commercial use. For example, the CBRS sharing regime builds on an Environmental Sensing Capability (ESC) and Spectrum Access System (SAS) to protect primarily a single incumbent, Navy radars, which only operates along the coasts. On the other hand, the 3.1 - 3.45 GHz band is home to a large number of different kinds of radars (airborne and land-mobile) deployed over the entire country and it will be extremely challenging to deploy ESC networks widely enough to monitor incumbent usage. These challenges are being considered by regulatory bodies like the National Telecommunications and Information Administration (NTIA) in the U.S.~\cite{ntia2023strategy} and Ofcom in the U.K.~\cite{ofcom20236ghz} as they develop spectrum-sharing strategies that will continue to meet the evolving needs of wireless connectivity.



\noindent{\bf{Our contributions:}} Motivated by the above discussion, we offer new insights in this paper through careful measurements and analyses of a real-world private network over CBRS in the U.S. that provides an indoor neutral-host to the customers of two MNOs with the goal of improving indoor cellular connectivity. Specifically, this work contributes the following:


\noindent $\bullet$
\textbf{First commercial private network measurements focused on indoor neutral-host deployments (\S\ref{sec:deployment}):} We believe our measurement campaign is the first to comprehensively measure a private neutral-host deployment. These measurements are carried out indoors and outdoors to quantify building and environmental losses.

\noindent $\bullet$
\textbf{Comparison of outdoor and indoor characterization of a neutral-host and mid-band 5G deployments (\S\ref{sec:coverage}):} We present a comprehensive analysis that includes deployment configurations, as well as heatmaps and distributions of the received signal strengths. By comparing indoor signal strengths from outdoor 5G deployments and outdoor reception of indoor CBRS deployments, we offer insights into signal propagation characteristics and provide a foundation for further investigation of building/environment loss.


\noindent $\bullet$
\textbf{Detailed insight into the performance of a neutral-host deployment (\S\ref{sec:dl_analysis},\S\ref{sec:ul_analysis}):} We show that the neutral-host network provides better indoor performance compared to MNO deployed 5G mid-band, in both downlink and uplink, due to the proximity and density of CBRS devices (CBSDs) deployed indoors. Moreover, the UE reports lower TX power indoors when connected to the neutral-host, and use of the neutral-host by indoor users creates additional capacity for the MNO to serve other customers by freeing up resource blocks.


These findings from real-world deployments will inform regulators as they make decisions regarding the future use of mid-band spectrum, including the 3.1 - 3.45~GHz, upper 6~GHz, and 7.125 - 8.4~GHz bands.

\begin{figure}[t]
    \begin{subfigure}{.24\textwidth}
    \includegraphics[width=\linewidth, height=3.5cm]{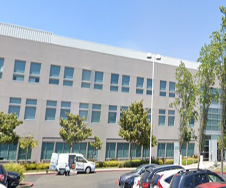}
    \caption{Overview of location}
    \label{fig:hc_env}
    \end{subfigure}
    \hfill
    \begin{subfigure}{.24\textwidth}
    \includegraphics[width=\linewidth, height=3.5cm]{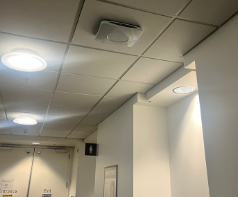}
    \caption{CBSD deployment}
    \label{fig:hc_antenna}
    \end{subfigure}
    \caption{Neutral-host deployment at the healthcare facility.}
    \label{fig:environment}
    \vspace{-.5em}
\end{figure}

\begin{table}
    \centering
    \caption{Deployment parameters.}
    \begin{tabular}{|C{10.5em}|C{6.5em}|C{6.5em}|}
        \hline
        \textbf{Parameters} & \multicolumn{2}{c|}{\textbf{Values}} \\
        \hline
        \hline
        \multicolumn{3}{|c|}{\textbf{Neutral-host deployment in CBRS}} \\
        \hline
        \# of deployed CBSD & \multicolumn{2}{C{13em}|}{1st floor: 5; 2nd floor: 5; \newline 3rd floor: 3} \\
        \hline
        \# of PCI & \multicolumn{2}{c|}{2 per CBSD} \\
        \hline
        Nature of Antenna & \multicolumn{2}{c|}{Omnidirectional} \\
        \hline
        TDD Config & \multicolumn{2}{c|}{DL: 4 subframes; UL: 4 subframes} \\
        \hline
        CBSD Tx Power & \multicolumn{2}{c|}{23 dBm} \\
        \hline
        Antenna Gain & \multicolumn{2}{c|}{5 dBi} \\
        \hline
        Band Number \& \newline Center Frequency & \multicolumn{2}{C{13em}|}{b48: \{3570, 3590, \newline 3670, 3690\}~MHz} \\
        \hline
        Bandwidth & \multicolumn{2}{c|}{20 MHz} \\
        \hline
        \hline
        \multicolumn{3}{|c|}{\textbf{Comparison MNO deployments}} \\
        \hline
        \textbf{} & \textbf{\att} & \textbf{\tmo} \\
        \hline
        Band Number \& \newline Center Frequency & n77: 3840~MHz & n41: \{2496, 2596\}~MHz \\
        \hline
        Bandwidth & 80 MHz & \{90, 100\} MHz \\
        \hline
        TDD Config & \multicolumn{2}{c|}{DL: 7 slots; UL: 2 slots} \\
        \hline
    \end{tabular}
    \label{table:exp_params}
    \vspace{-1em}
\end{table}


\section{Related Work}


The low-power-indoor (LPI) ruling in the 6~GHz band~\cite{FCC4} fosters effective spectrum reuse while protecting incumbents fixed link operations by simply limiting the transmit power of indoor devices. Extensive measurements of dense LPI Access Points (APs) at the University of Michigan and the University of Notre Dame demonstrated that most indoor signals are well contained within buildings~\cite{dogan2023evaluating,dogan2023indoor}. While a small number of APs near windows show high outdoor RSSI (Received Signal Strength Indicator) levels, this occurs in limited locations and do not pose a significant probability of harmful interference to incumbents which are primarily deployed outdoors with narrow, directional antennas on high towers.

The CBRS band has also demonstrated a successful shared spectrum framework with categorized access for low-power indoor (Cat A) and medium-power outdoor (Cat B) operations. This approach effectively protects outdoor incumbent naval and satellite operations while enabling indoor private networks for critical applications. In~\cite{sathya2023warehouse}, the author analyzed the commercial indoor private network deployment of 23 CBSDs in a warehouse environment. The study highlights the reliability and coexistence potential of CBRS, which is enabled by the SAS. Another successful commercial indoor deployment within an enterprise setting also shows efficient and reliable connectivity without compromising protection for incumbent users~\cite{sathya2022comparative}.

Looking ahead, the US is exploring spectrum sharing in the 3.1 - 3.45~GHz band currently used by the Department of Defense. Unlike the static nature of ship radars in CBRS, airborne transmissions in this band require a more dynamic sharing approach. The insights gained from our measurement studies in CBRS will be invaluable for future spectrum policy and regulation developments in this band.

\begin{table}
    \centering
    \caption{Captured LTE and NR parameters from Qualipoc.}
    \label{tab:qp_params}
    \begin{tabular}{|L{5em}|L{23em}|}
        \hline
        \textbf{Parameter} & \textbf{Description} \\
        \hline
        \hline
        Latitude, Longitude & UE’s geographic coordinates calculated by the Android API \\
        \hline
        PCI & Physical Cell Identifier \\
        \hline
        RSRP/\newline RSRQ & Signal strength values. For NR, RSRP/RSRQ indicates measurements from the 5G Synchronization Signal (SS) block. \\
        \hline
        DL/UL ARFCN & Downlink/Uplink Absolute Radio Frequency Channel Number, \ie center frequency. \\
        \hline
        PDSCH/\newline PUSCH Throughput & Throughput values recorded on the Physical Downlink Shared Channel and Physical Uplink Shared Channel, \ie PHY-layer downlink and uplink throughput. \\
        \hline
        DL/UL MCS & Modulation and Coding Scheme utilized in DL and UL channels.  \\
        \hline
        DL BLER & Block Error Rate for DL channel. \\
        \hline
        DL/UL RB & Number of allocated Resource Block per Subframe (LTE) or Slot (NR) in DL and UL.  \\
        \hline
        DL MIMO Layer & Number of utilized MIMO layer in DL transmissions. This parameter is only available in 5G. \\
        \hline
        Pathloss & Pathloss as calculated by the UE, using Reference Signal TX power and RSRP.  \\
        \hline
        PUSCH TX power & Uplink TX power used by the UE. \\
        \hline
    \end{tabular}
    \vspace{-1em}
\end{table}

\section{Deployment, Methodology and Tools} \label{sec:deployment}

The measurements analyzed in this work were collected on November 25, 2024, at a healthcare facility where a neutral-host CBRS network is deployed indoors using GAA. Fig.~\ref{fig:environment} presents an overview of the measurement location and the deployed CBSDs, while Table~\ref{table:exp_params} summarizes the parameters of the CBRS deployment and the two MNOs, \att and \tmo that are supported by the neutral-host network. The CBSDs are installed on the ceiling facing down on three different floors. The optimal placement of the CBSDs was determined using IBWAVE RF planning~\cite{ibwave}, considering building materials such as concrete, wood, and glass walls. All CBSDs utilize 20 MHz channels within the CBRS band (band number b48 / 3.55-3.7 GHz), as allocated by the SAS. These channels employ Time Division Duplexing (TDD) with a balanced configuration of 4 downlink (DL), 4 uplink (UL) and 2 special subframes per 10 ms radio frame. Up to two b48 channels can be aggregated in the network.
\att has deployed a 5G Non-Standalone (NSA) network, aggregating LTE and NR channels, while \tmo employs a 5G Standalone (SA) configuration, aggregating up to three NR channels. To facilitate comparison with the CBRS b48 band, we focus our analysis on the n77 (\att) and n41 (\tmo) channels: one 80 MHz wide n77 channel and two n41 channels with 90 and 100~MHz bandwidths are used. Both n41 and n77 deploy a TDD configuration with 7~DL, 2~UL, and 1~special slots every 5 subframes (0.5 ms).

The measurements were conducted by walking indoors on all three floors of the facility, and outdoors around the parking lot.  A Samsung S23+ smartphone equipped with the Qualipoc app~\cite{qualipoc} was used to collect low-level data from the cellular modem via the Qualcomm \textit{Diag} (diagnostic) interface. Table~\ref{tab:qp_params} presents a subset of the LTE and NR parameters used in our analysis. Each parameter is immediately captured upon the availability of a new value, resulting in varying capture intervals with the shortest of around 100 ms.
The Qualipoc app also collected DL and UL throughput performance by initiating HTTP traffic, alternating every 5 seconds between an HTTP GET method to download a 2~GB file hosted at GitHub.com and a POST method to upload a 1~GB file to HTTPbin.org.
\chk{
The Qualipoc app collects raw modem logs which further processed by the SmartAnalytics back-end and exported as CSV files. These CSV files are then used for data analysis and visualization.
}
After filtering based on the bands shown in Table~\ref{table:exp_params}, we collected a total of 31,720 data points over approximately 200 minutes of measurements.



\begin{figure*}
    \begin{subfigure}{.32\textwidth}
    \includegraphics[width=\linewidth]{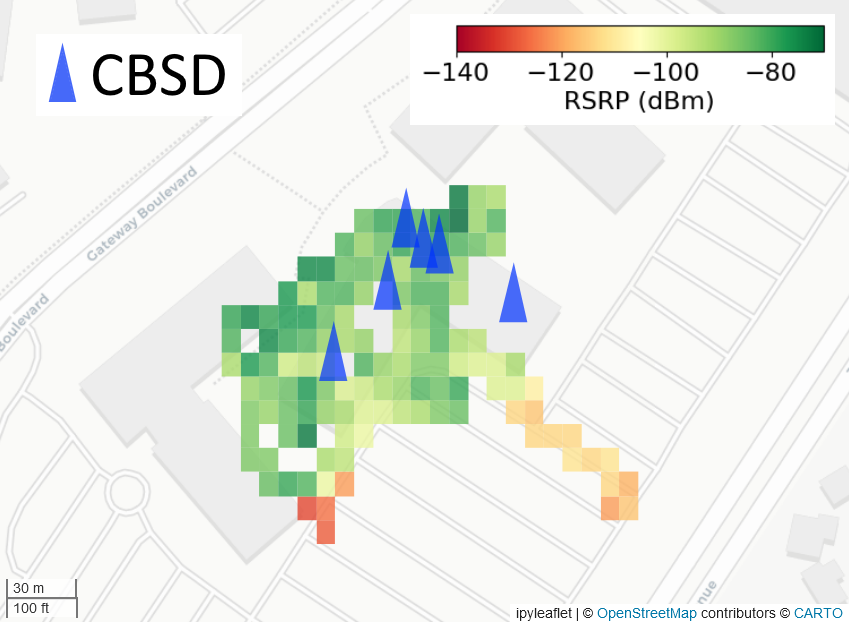}
    \caption{Neutral-host b48, all PCIs}
    \label{fig:heatmap_rsrp_cln}
    \end{subfigure}
    \hfill
    \begin{subfigure}{.32\textwidth}
    \includegraphics[width=\linewidth]{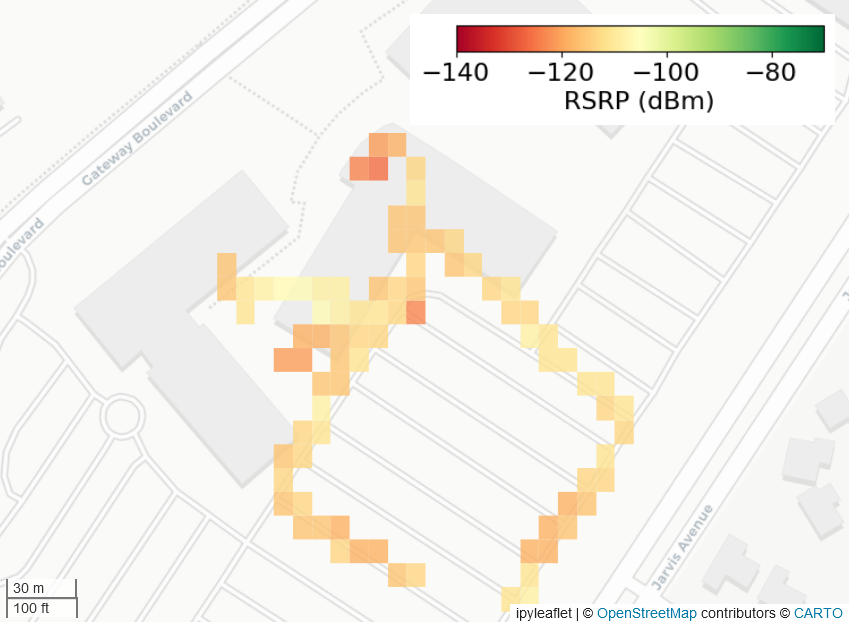}
    \caption{\att n77, all PCIs}
    \label{fig:heatmap_rsrp_att}
    \end{subfigure}
    \hfill
    \begin{subfigure}{.32\textwidth}
    \includegraphics[width=\linewidth]{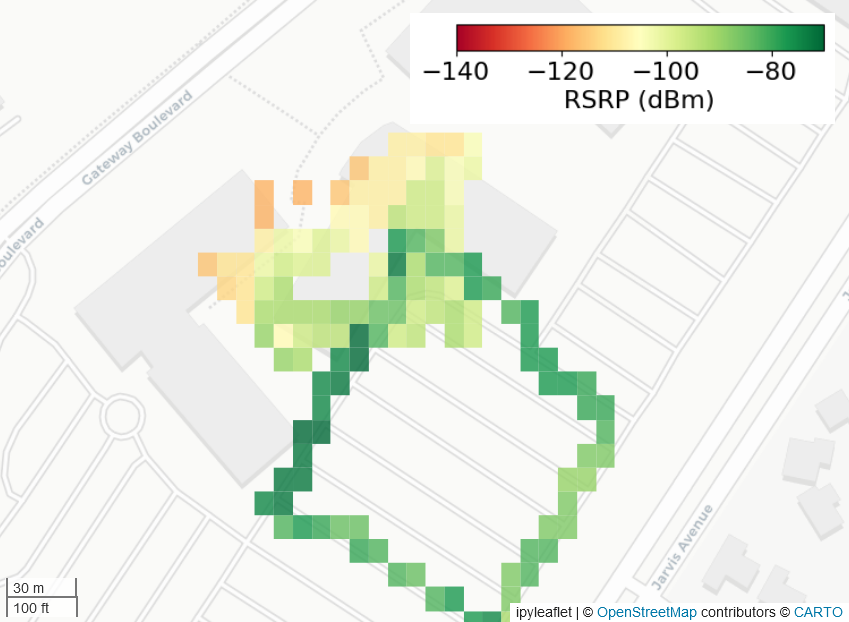}
    \caption{\tmo n41, all PCIs}
    \label{fig:heatmap_rsrp_tmo}
    \end{subfigure}
    \caption{RSRP heatmaps of neutral-host and MNO's 5G deployments.}
    \label{fig:heatmap_rsrp}
    \vspace{-1.5em}
\end{figure*}

\section{Results \& Discussions}

\subsection{Comparison of Indoor and Outdoor Coverage}\label{sec:coverage}


To provide a spatial overview of coverage, Fig.~\ref{fig:heatmap_rsrp} presents RSRP heatmaps which are generated by binning RSRP values over 10 m$^2$ bins and generating representative values per bin by averaging RSRP values in the linear domain.
Fig.~\ref{fig:heatmap_rsrp_cln} displays the RSRP heatmap of the neutral-host network, captured with the UE equipped with either \att or \tmo SIMs. (blue triangles indicate CBSD positions). The heatmap reveals robust indoor coverage but weaker outdoor coverage, with no neutral-host signal detected in the southern section of the parking lot. It should be noted that measurements were taken while walking around the entire parking lot, but measurable signal levels were obtained only in the regions shown in Fig.~\ref{fig:heatmap_rsrp_cln}.
Comparing with MNO deployments, Fig.~\ref{fig:heatmap_rsrp_att} illustrates weak n77 coverage both indoors and outdoors, while Fig.~\ref{fig:heatmap_rsrp_tmo} shows strong outdoor coverage but weak indoor coverage for n41.

\begin{figure}
    \includegraphics[width=\linewidth]{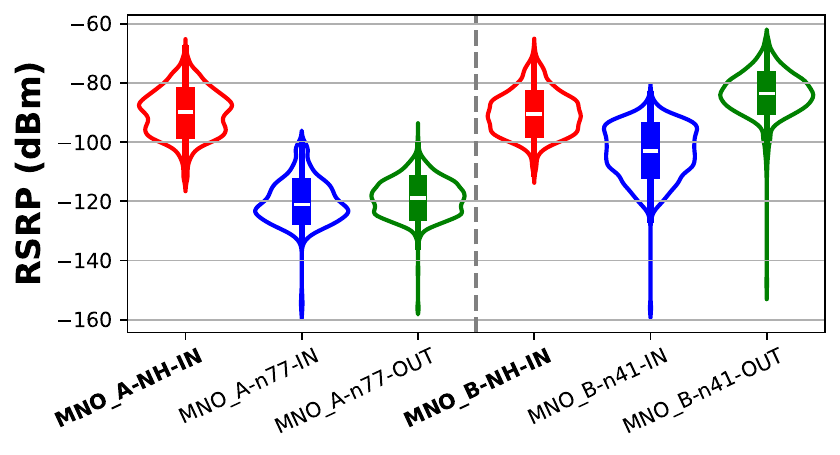}
    \caption{Comparison of RSRP for all observed PCIs.}
    \label{fig:violin_rsrp_new1}
    \vspace{-1em}
\end{figure}

\begin{figure}
    \includegraphics[width=\linewidth]{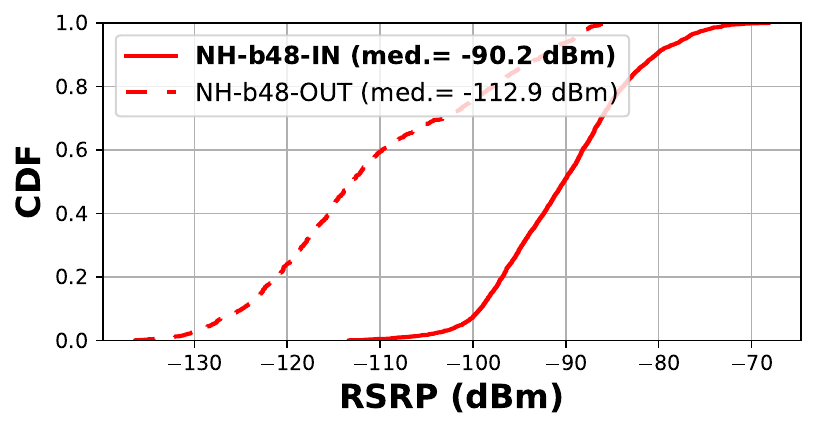}
    \caption{Comparison of neutral-host RSRP observed indoors and outdoors.}
    \label{fig:cdf_rsrp_NH_new1}
    \vspace{-1.5em}
\end{figure}

\begin{figure}
    \includegraphics[width=\linewidth]{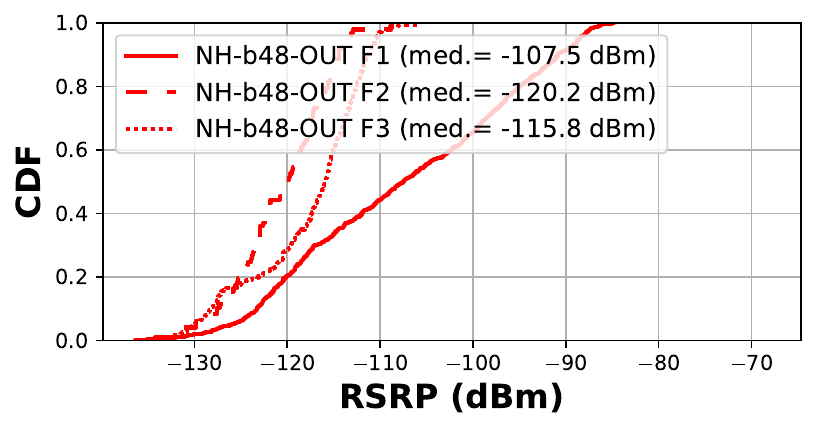}
    \caption{Comparison of neutral-host's outdoor RSRP based on the floor of the transmitting CBSD.}
    \label{fig:cdf_rsrp_NH_floors}
    \vspace{-1em}
\end{figure}

Fig.~\ref{fig:violin_rsrp_new1} corroborates the low \att reception indoors and outdoors, and the better outdoor reception of \tmo compared to its indoors. Additionally, we compare indoor signal reception between the neutral-host and the MNOs.
When subscribed to \att, the neutral-host network served indoor UE with a median RSRP of -89~dBm (\textbf{MNO\_A-NH-IN} in the plot). Similarly, the median of neutral-host's RSRP is -90~dBm when UE is subscribed to \tmo (\textbf{MNO\_B-NH-IN}).
These indoor RSRP values surpasses that of both MNOs' indoor RSRPs and is even comparable to \tmo's outdoor reception.
This strong indoor coverage from the neutral-host network can be attributed to the high density of deployed CBSDs: there are 13 CBSDs across the 3 floors, each with 2 PCIs, resulting in a total of 26 PCIs.
However, we observe that the signal attenuates rapidly, with a 22 dB reduction in the median when the UE is outdoors, as shown in Fig.~\ref{fig:cdf_rsrp_NH_new1}. This demonstrates significant penetration loss, which can enable effective sharing with outdoor incumbents. 
Fig.~\ref{fig:cdf_rsrp_NH_floors} further analyzes outdoor RSRP based on the floor on which the transmitting CBSD is installed (F1: 1st Floor, F2: 2nd Floor, F3: 3rd Floor). As expected, we observe the highest median RSRP outdoors for the CBSDs installed on the first floor. Note that the CBSDs deployed on the third floor have a better line-of-sight to the outdoor parking lot, resulting in a higher RSRP than those on the second floor.

\begin{figure}
    \includegraphics[width=\linewidth]{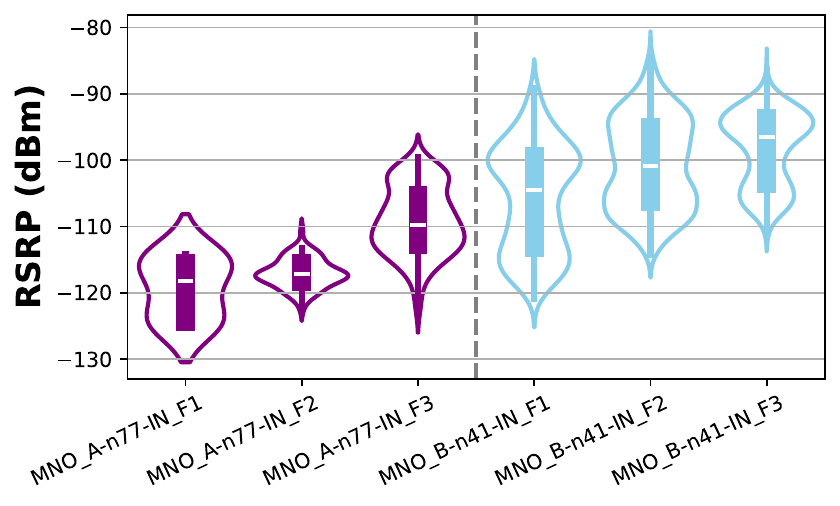}
    \caption{Comparison of MNO's indoor RSRP on various floors.}
    \label{fig:violin_rsrp_MNO_floors}
    \vspace{-1em}
\end{figure}

Table~\ref{tab:pci-stats} summarizes the availability of MNO PCIs indoors and outdoors. Overall, we observe a higher number of \att PCIs compared to \tmo, despite \tmo exhibiting stronger RSRP distributions. This contradicts the expectation that denser deployments (more PCIs) correlate with better signal reception. We attribute this discrepancy to the limited measurement footprint around the healthcare facility. A city-wide measurement campaign would likely reveal a clearer correlation between PCI density and signal quality.
Analyzing floor-by-floor, we observe an increasing number of PCIs from the first floor (IN F1) to the third floor (IN F3) for both \att and \tmo.
Furthermore, the set of PCIs observed on lower floors consistently forms a subset of those observed on higher floors.
However, a discrepancy arises when we observe certain PCIs indoors, but not outdoors, and vice versa. For instance, we observe one \tmo PCI outdoors but not indoors, and four PCIs indoors but not outdoors. This disparity is again attributed to the limitations of our measurement footprint. The indoor PCIs not observed outdoors were primarily captured on the north side of the building or on higher floors, while outdoor PCIs not observed indoors were captured on the south side of the parking lot.
Only a handful of PCIs are connected to our UE outdoors, with only one PCI connected indoors.
When analyzing both MNOs' indoor RSRP on different floors, we observe positive correlation between median RSRP and the floor, as shown on Fig.~\ref{fig:violin_rsrp_MNO_floors}.

\begin{table}[t]
\centering
\caption{Observed MNOs PCIs}
\label{tab:pci-stats}
\resizebox{\columnwidth}{!}{%
\begin{tabular}{|c|c|C{8em}|C{6.5em}|}
\hline
\multicolumn{2}{|c|}{\textbf{Category and Location}} &
  \textbf{MNO-A} &
  \textbf{MNO-B} \\ \hline
\hline
&
  \textbf{IN F1} &
  4 PCIs &
  3 PCIs \\ \cline{2-4}
\textbf{All (connected}&
  \textbf{IN F2} &
  9 PCIs &
  4 PCIs \\ \cline{2-4} 
\textbf{and neighboring)}&
  \textbf{IN F3} &
  22 PCIs &
  6 PCIs \\ \cline{2-4} 
&
  \textbf{OUT} &
  31 PCIs &
  3 PCIs \\ \hline
\hline
\multirow{2}{*}{\textbf{Connected}} &
  \textbf{IN All Floors} &
  PCI 784 &
  PCI 420 \\ \cline{2-4} 
&
  \textbf{OUT} &
  PCI 371, 737, 784 &
  PCI 373, 420 \\ \hline
\end{tabular}%
}
\end{table}

\subsection{Downlink performance comparison}\label{sec:dl_analysis}

To compare indoor downlink performance between the neutral-hosts and the MNO, we analyze data associated with PCIs connected to the UE indoors. We observe a substantial difference in the number of PCIs: all 26 neutral-host PCIs were connected indoors, while only PCI 784 in n77 and PCI 420 in n41 served \att and \tmo, respectively, as shown in Table~\ref{tab:pci-stats}. This highlights the density of neutral-host deployment which is focused on serving indoor users. Additionally, we include the outdoor performance of the representative PCIs in the plots \textit{only} to compare it with the indoor counterparts.

\begin{figure}
    \begin{subfigure}{.24\textwidth}
    \includegraphics[width=\linewidth]{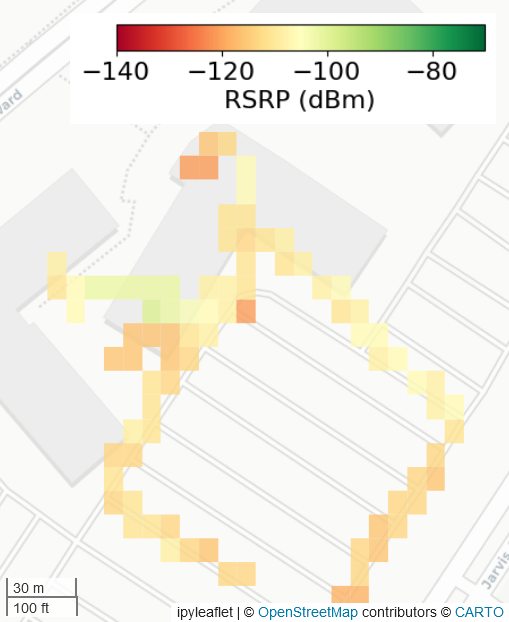}
    \caption{\att n77 PCI 784}
    \label{fig:heatmap_rsrp_att_pci}
    \end{subfigure}
    \hfill
    \begin{subfigure}{.24\textwidth}
    \includegraphics[width=\linewidth]{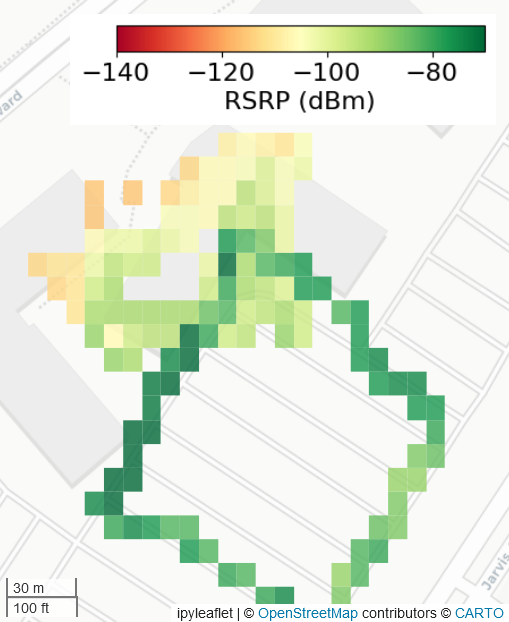}
    \caption{\tmo n41 PCI 430}
    \label{fig:heatmap_rsrp_tmo_pci}
    \end{subfigure}
    \caption{RSRP heatmaps for each MNO's representative PCI.}
    \label{fig:heatmap_rsrp_pci}
    \vspace{-1em}
\end{figure}

\begin{figure}
    \includegraphics[width=\linewidth]{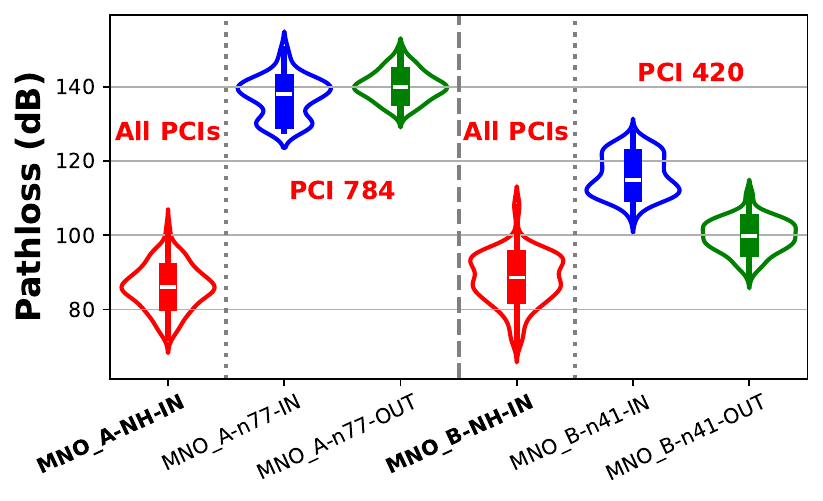}
    \caption{Comparison of pathloss between all neutral-host PCIs and each MNO's representative PCI.}
    \label{fig:violin_pathloss_new1}
    \vspace{-1em}
\end{figure}

Fig.~\ref{fig:heatmap_rsrp_pci} illustrates the coverage of each representative PCI. We observe a 7~dB higher median RSRP for \att's PCI 784 compared to all PCIs that utilizes n77, while \tmo PCI 430's median RSRP is 1~dB higher than all n41 PCIs. This indicates that \tmo's coverage is relatively uniform outdoors.
Fig.~\ref{fig:violin_pathloss_new1} compares pathloss, calculated by the UE using reference signal TX power and RSRP.  We observe a definite improvement in indoor pathloss when the UE is served by the neutral-host. Again, this can be attributed to the density of the neutral-host deployment. On the other hand, the pathloss of \att PCI 784 is similar both indoors and outdoors, while \tmo PCI 420 shows better outdoor pathloss distribution. 

\begin{figure}
    \includegraphics[width=\linewidth]{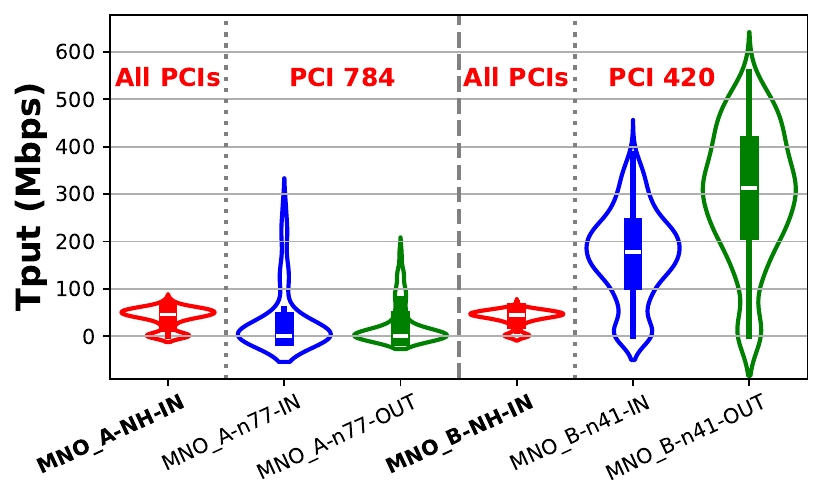}
    \caption{Comparison of DL throughput between all neutral-host PCIs and each MNO's representative PCI.}
    \label{fig:violin_dl_tput_new1}
\end{figure}

\begin{figure}
    \includegraphics[width=\linewidth]{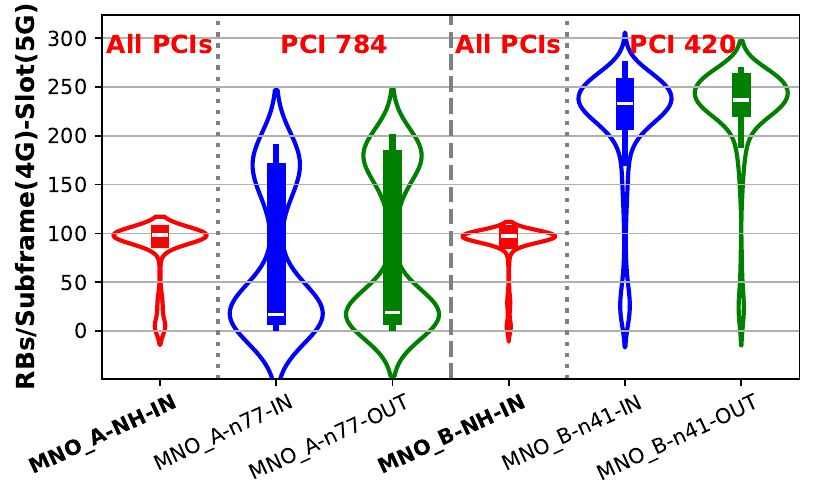}
    \caption{Comparison of DL RB allocation per subframe (4G) or slot (5G) between all neutral-host PCIs and each MNO's representative PCI.}
    \label{fig:violin_dl_rb_per_unit_new1}
    \vspace{-1.5em}
\end{figure}

Fig.~\ref{fig:violin_dl_tput_new1} compares downlink throughput as reported by the UE. Notably, \tmo's representative PCI 420 exhibits significantly higher indoor and outdoor downlink throughput than both \att and the neutral-host due to its wider bandwidth (RB allocation) and strong signal reception.
Consistent with the previous RSRP and pathloss analysis, we observe poor DL throughput for \att both indoors and outdoors, and better performance for \tmo outdoors compared to its indoors.
The median DL throughput achieved by the neutral-host's b48 channel while subscribed to \att and \tmo are around 45~Mbps. These values are considerably higher compared to \att with a median of 84~kbps, \ie a 535$\times$ increase. 
Fig.~\ref{fig:violin_dl_rb_per_unit_new1} confirms the DL throughput analysis: median of RB allocation for neutral-host is higher than \att n77 indoors. RB allocations in \tmo n41 is also significantly higher in both environments: we observe a high median allocation of 233 RBs indoors in \tmo's channels, which can be freed by offloading the UE to the neutral-host network.
Moreover, we observe a consistent utilization of two MIMO layers in the neutral-host's b48 channels enabled by the dense deployment.

\begin{figure}
    \includegraphics[width=\linewidth]{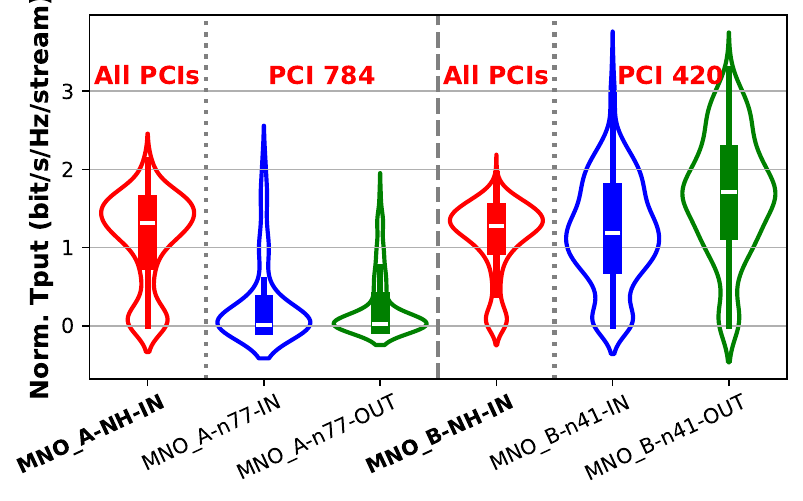}
    \caption{Comparison of normalized DL throughput between all neutral-host PCIs and each MNO's representative PCI.}
    \label{fig:violin_dl_tput_norm_new1}
\end{figure}

To further analyze channel utilization, a spectral efficiency metric is defined by normalizing throughput with respect to bandwidth and the number of MIMO layers:
$Tput_{norm} = Tput_{bps} / (N_{RB} \times SCS_{Hz} \times 12) / N_{layer})$,
where $Tput_{bps}$ is throughput in bits/second, $N_{RB}$ is the average number of resource blocks (RBs) allocated to the UE over one second, $SCS_{Hz}$ is the subcarrier spacing (SCS) in Hz, and $N_{layer}$ is the number of MIMO layers used.  Instantaneous bandwidth usage is determined by multiplying $N_{RB}$ by $SCS_{Hz}$ and 12, as each RB contains 12 subcarriers. Normalizing by RB count effectively accounts for differences in TDD configurations across the channels.
As computed above, Fig.~\ref{fig:violin_dl_tput_norm_new1} compares the normalized throughput performance of the neutral-host, \att, and \tmo. The neutral-host network shows a median of 1.3~bit/s/Hz/stream, indoors while subscribed to either \att or \tmo. These values are higher compared to \att's representative PCI (indoor median: 0.01~bit/s/Hz/stream) but comparable to \tmo's (indoor median: 1.2~bit/s/Hz/stream).

\begin{figure}
    \includegraphics[width=\linewidth]{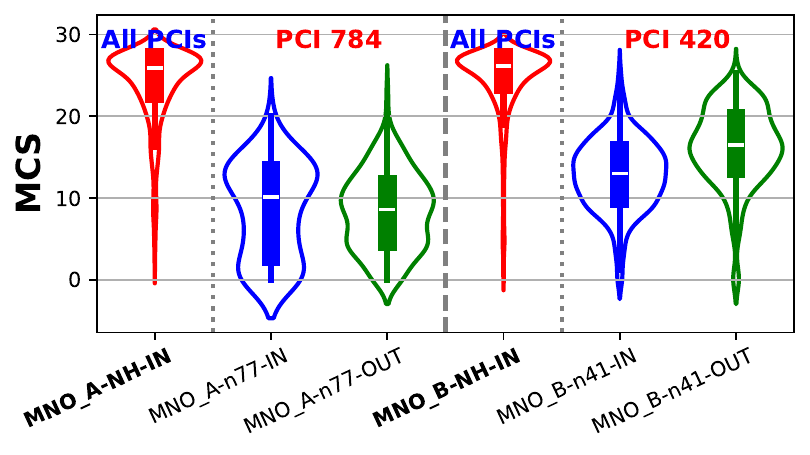}
    \caption{Comparison of DL MCS between all neutral-host PCIs and each MNO's representative PCI.}
    \label{fig:violin_dl_mcs_new1}
    \vspace{-1.5em}
\end{figure}

We observed the usage of MCS Table 2 as described in Table 5.1.3.1-2 of \cite{3gpp17-38214} on both 4G neutral-host and 5G MNOs, as indicated by the RRC messages. Therefore, direct comparison of MCS values across these different radio access technologies is valid.
Fig.~\ref{fig:violin_dl_mcs_new1} demonstrates that the neutral-host deployment consistently outperforms both \att and \tmo indoors in terms of higher MCS values. This is again attributed to the dense indoor deployment of the neutral-host network.  We observe a median MCS of 26 when the UE is served indoors by the neutral-host, regardless of whether the UE is subscribed to \att or \tmo. This is higher than the median MCS values for \att and \tmo's representative PCIs, which are 10 and 13, respectively.
Consistent with prior analyses, \att shows similar median MCS values indoors and outdoors, while \tmo exhibits a lower median MCS indoors.
Overall, all of our DL performance metrics show weak \att performance on both environments, which is improved by the neutral-host network.

\subsection{Uplink performance comparison}\label{sec:ul_analysis}



\begin{figure}
    \includegraphics[width=\linewidth]{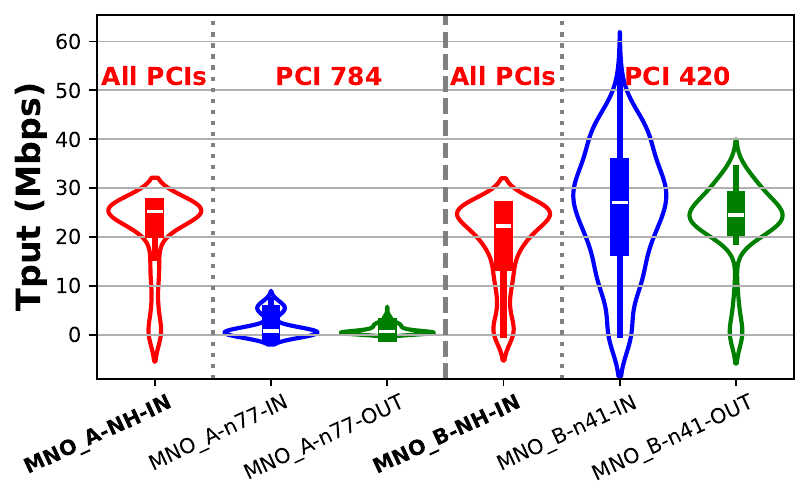}
    \caption{Comparison of UL throughput between all neutral-host PCIs and each MNO's representative PCI.}
    \label{fig:violin_ul_tput_new1}
\end{figure}

\begin{figure}
    \includegraphics[width=\linewidth]{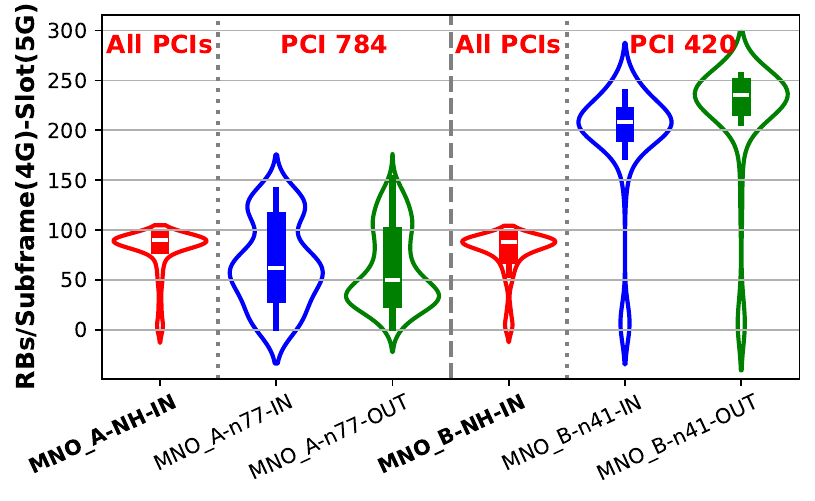}
    \caption{Comparison of UL RB allocation per subframe (4G) or slot (5G) between all neutral-host PCIs and each MNO's representative PCI.}
    \label{fig:violin_ul_rb_per_unit_new1}
    \vspace{-1em}
\end{figure}

Similar to the DL performance analysis, we filter MNO's data based on the representative PCIs: 784 for \att and 420 for \tmo.
Fig.~\ref{fig:violin_ul_tput_new1} shows that the neutral-host's b48 achieves 25~Mbps indoors when the UE is subscribed to \att, higher than \att n77  (median: 757~kbps). On the other hand, when subscribed to \tmo, the neutral-host achieved a median of 22~Mbps which is slightly lower than \tmo n41 (median: 27~Mbps).
\att's uplink throughput remains consistently low indoors and outdoors, which further reinforces its overall weak network performance at the site. Using the worse-performing \att as a baseline, we observe 33$\times$ increase in UL throughput.
Fig.~\ref{fig:violin_ul_rb_per_unit_new1} confirms the uplink throughput analysis, particularly regarding the distribution of MNOs' uplink RB allocation indoors and outdoors. Similar to the downlink RB observation, we can infer that resources allocated to the MNOs for indoor uplink transmission (up to a median of 208 RBs) can be freed by offloading users to the neutral-host network.

\begin{figure}
    \includegraphics[width=\linewidth]{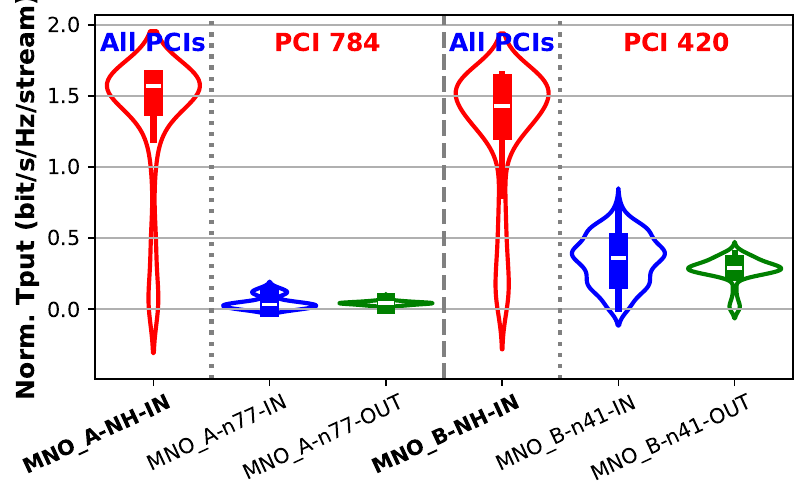}
    \caption{Comparison of normalized UL throughput between all neutral-host PCIs and each MNO's representative PCI.}
    \label{fig:violin_ul_norm_tput_new1}
\end{figure}

\begin{figure}
    \includegraphics[width=\linewidth]{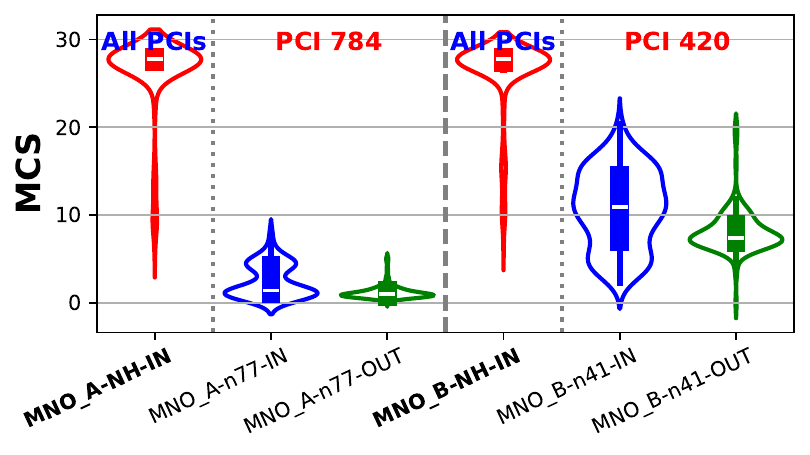}
    \caption{Comparison of UL MCS between all neutral-host PCIs and each MNO's representative PCI.}
    \label{fig:violin_ul_mcs_new1}
    \vspace{-1em}
\end{figure}

The Qualipoc app does not capture the number of MIMO layers used for uplink in 4G. However, we confirmed that the CBSD supports only one MIMO layer in UL, and we use this value to calculate the neutral-host's normalized UL throughput.
As shown in Fig.~\ref{fig:violin_ul_norm_tput_new1}, neutral-host's b48 outperforms both \att and \tmo indoors, particularly with the highest median of 1.5 and 1.4~bit/s/Hz/stream on either \att and \tmo SIMs, respectively. \att exhibits a median of 0.03 and 0.04~bit/s/Hz/stream indoors and outdoors, respectively, while \tmo achieved 0.35 and 0.29~bit/s/Hz/stream indoors and outdoors, respectively. 
Fig.~\ref{fig:violin_ul_mcs_new1} further corroborates the normalized UL throughput observations, showing that the neutral-host network assigns significantly higher MCS values to the UE when indoors. This can be attributed to the favorable channel conditions provided by the dense CBSDs deployment.

\begin{figure}
    \includegraphics[width=\linewidth]{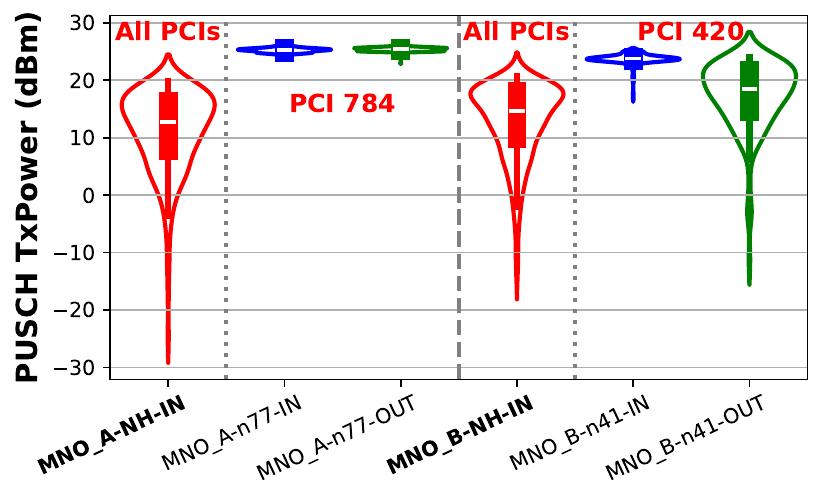}
    \caption{Comparison of PUSCH Tx power between all neutral-host PCIs and each MNO's representative PCI.}
    \label{fig:violin_ul_txpow_new1}
    \vspace{-1em}
\end{figure}

Finally, we compare the TX power utilized by the UE to transmit PUSCH. Fig.~\ref{fig:violin_ul_txpow_new1} shows median TX power values of 13~dBm when the UE utilizes the neutral-host's b48 channels while subscribed to \att, and 15~dBm while subscribed to \tmo. These values are considerably lower than those for \att n77 (median: 25~dBm) and \tmo n41 (median: 23~dBm) indoors.  \att uses the same TX power indoors and outdoors, as evidenced by the consistent median of 25~dBm.  \tmo, on the other hand, exhibits the expected behavior, with a median TX power 5 dB lower outdoors compared to indoors. Across all metrics, our observations indicate considerable enhancements in UL performance when the UE is served by the neutral-host network, along with reduced UE power consumption.

\subsection{Discussions}

This measurement campaign, conducted within a healthcare facility, provides a snapshot on the capabilities of a neutral-host deployment. However, due to its singular environment, it may not represent the full spectrum of potential deployment scenarios, such as manufacturing or warehouse complexes. The healthcare facility, which is a typical three-story concrete building, exhibited a median building loss of 22 dB in the CBRS spectrum. In contrast, \att (n77) showed only a 1 dB loss, likely due to its already weak outdoor reception (a median RSRP of -120 dBm), while \tmo (n41) demonstrated a median loss of 19 dB. These results suggest effective indoor-outdoor isolation within this environment, though building loss can vary based on materials and MNO base station proximity.

The indoor performance gains, including downlink and uplink throughput and reduced uplink transmit power, are contingent upon CBSD deployment planning and existing MNO coverage. The healthcare facility's 13 CBSDs were strategically deployed, ensuring comprehensive indoor coverage. Among the MNOs, only \att experienced significant throughput gains, attributed to its suboptimal indoor and outdoor coverage. While \tmo, already delivered strong performance with its robust n41 channels (up to 100 MHz bandwidth), exceeding the 20 MHz channel width of the neutral-host CBRS. This exemplify the trade-off of neutral-host networks: there is no need for neutral-host deployments if macro deployments already dense enough to cover indoors, or if the spectrum capacity hasn't reached.

Despite the limitations of a single measurement location, these findings demonstrate the potential of neutral-host systems for indoor-outdoor spectrum sharing in mid-band frequencies. Further measurements across diverse environments and on a longer time scale are necessary to validate and expand upon these initial observations.

\section{Conclusions \& future research}

The paper presented detailed measurements and quantitative analyses of a real-world neutral-host deployment over the CBRS band, serving two MNOs with distinct macro-cell deployments.
\chk{While this measurement campaign is limited to one deployment location, we believe this is the first that comprehensively measure a neutral-host deployment.} 
The neutral-host solution significantly improved both downlink and uplink throughput for the MNO with initially lower performance: up to 535$\times$ and 33$\times$ at the median, respectively. Both MNOs also observed benefits including reduced user device power consumption during uplink transmissions (up to 12~dB reduction) and increased network capacity due to freed resource blocks (up to 233 RBs/slot). These resources can be reallocated to enhance service for outdoor users without alternative connectivity options.
Furthermore, the median outdoor signal strength due to the indoor deployment was 22~dB lower than indoors, 
which highlights the feasibility of low-power indoor deployments within shared spectrum bands. Such deployments offer a straightforward approach to spectrum sharing, improving indoor connectivity without disrupting outdoor, perhaps federal, incumbent operations.
At the same time, offloading indoor traffic to shared CBRS spectrum increases capacity on existing high-power, exclusively licensed MNO bands. This approach eliminates the need to relocate incumbents or develop complex sharing mechanisms for new exclusive licenses, ultimately increasing overall spectrum efficiency.

Our future work in this area will compare performance over neutral-host with indoor Wi-Fi and extend the analysis to quantifying latency improvements. \chk{Moreover, we plan to measure varying deployment environment, \eg office spaces, industrial/manufacturing plants, and large warehouses.}

\section*{\centering{Acknowledgements}}

This work is supported by NSF grants CNS-2229387, CNS-2346413, and AST-2132700.

\bibliographystyle{IEEEtran}
\bibliography{main}

\end{document}